\documentclass{IEEEcsmag}

\usepackage[colorlinks,urlcolor=blue,linkcolor=blue,citecolor=blue]{hyperref}
\expandafter\def\expandafter\UrlBreaks\expandafter{\UrlBreaks\do\/\do\*\do\-\do\~\do\'\do\"\do\-}
\usepackage{upmath,color}
\usepackage{listings}
\usepackage{xurl}
\usepackage[normalem]{ulem}
\usepackage{doi}

\newcommand{\addtext}[1]{#1}

\lstdefinestyle{yaml}{
     basicstyle=\color{blue}\footnotesize,
     rulecolor=\color{black},
     string=[s]{'}{'},
     stringstyle=\color{blue},
     comment=[l]{:},
     commentstyle=\color{black},
     morecomment=[l]{-},
    numbers=left,                    
    numbersep=5pt,
    captionpos=t
 }

\jvol{jvol}
\jnum{jnum}
\paper{paper}
\jmonth{Month}
\jname{IEEE CiSE}
\jtitle{Do Research Software Engineers and Software Engineering Researchers Speak the Same Language?}
\pubyear{2025}

\begin{document}

\sptitle{Article Type: Description  (see below for more detail)}

\title{Do Research Software Engineers and Software Engineering Researchers Speak the Same Language?}

\author{Timo Kehrer}
\affil{University of Bern, CH-3006 Bern, Switzerland}

\author{Robert Haines}
\affil{University of Manchester, Manchester, M13 9PL, UK}

\author{Guido Juckeland}
\affil{Helmholtz-Zentrum Dresden Rossendorf, 01328 Dresden, Germany}

\author{Shurui Zhou}
\affil{University of Toronto, Toronto, ON, M5S 3G4, CA}

\author{David E.~Bernholdt}
\affil{Oak Ridge National Laboratory, Oak Ridge, TN, 37831, USA}

\markboth{THEME/FEATURE/DEPARTMENT}{THEME/FEATURE/DEPARTMENT}

\begin{abstract}
Anecdotal evidence suggests that Research Software Engineers (RSEs) and Software Engineering Researchers (SERs) often use different terminologies for similar concepts, creating communication challenges. To better understand these divergences, we have started investigating how SE fundamentals from the SER community are interpreted within the RSE community, identifying aligned concepts, knowledge gaps, and areas for potential adaptation. Our preliminary findings reveal opportunities for mutual learning and collaboration, and our systematic methodology for terminology mapping provides a foundation for a crowd-sourced extension and validation in the future.
\end{abstract} 
\maketitle

\begin{figure}[!b]
\noindent\fbox{%
    \parbox{0.95\linewidth}{
This manuscript has been authored in part by UT-Battelle, LLC, under contract DE-AC05-00OR22725 with the US Department of Energy (DOE). The US government retains and the publisher, by accepting the work for publication, acknowledges that the US government retains a non-exclusive, paid-up, irrevocable, world-wide license to publish or reproduce the submitted manuscript version of this work, or allow others to do so, for US government purposes. DOE will provide public access to these results of federally sponsored research in accordance with the DOE Public Access Plan (\url{https://energy.gov/doe-public-access-plan}).
}
}
\end{figure}

\section{Motivation and Goals}
\label{sec:motivation}

In science and technology, communities often engage predominantly within their own circles, which fosters the development of unique terminologies. As a result, members of different communities may find their interpretations of key terms diverging, often leading to unintentional miscommunication. Thus, establishing a common ``language'' that facilitates effective communication is key for mutual understanding and bridging the gap between communities. This is particularly relevant for the two major communities addressed by the Dagstuhl Seminar 24161 ``Research Software Engineering: Bridging Knowledge Gaps''~\cite{druskat_et_al:DagRep.14.4.42}. On the one hand, there is the traditional software engineering research (SER) community~\cite{shaw2002makes}, which systematically investigates software engineering (SE) methods, tools, processes, and practices to which we collectively refer to as \emph{SE fundamentals}. On the other hand, there is the much younger but rapidly growing research software engineering (RSE) community~\cite{cohen2020four}, which has developed its informal understanding of SE fundamentals through the practical experience of creating research software (RS).  Anecdotally, few research software developers have any formal educational experience with software engineering practices, instead acquiring their understanding of SE on an as-needed basis, often from colleagues or second-hand materials rather than textbooks or the SER literature.

While both RSE and SER share the goal of developing high-quality software, anecdotal evidence suggests that they sometimes use different terminologies for the same concepts. Moreover, the RSE community may lack awareness of established practices from SER, preventing them from fully leveraging those practices in their development processes. Conversely, some best practices identified by SER may not be directly applicable to the domain-specific context of RSE. This highlights a notable gap between the two communities.
\addtext{The SER community may not even be aware of best practices that emerged within RSE, but has just recently started to treat RSE as a research subject~\cite{5337646,hasselbring2024researchsoftwarecategories,10.1145/3685265}.}

In order to better understand the situation, and as a precursor to efforts to close gaps between the RSE and SER communities, our Dagstuhl Seminar working group decided to delve deeper by examining the mapping between the terms used in the two groups.
While the overarching goal and potential impact of such terminology mapping was undisputed among the members of the working group, it became evident that identifying a comprehensive and mutually accepted collection of SE fundamentals is a labor-intensive task, and that mapping these fundamentals is subject to intense debate, often influenced by the diverse professional backgrounds of the group members. Given the limited time available during the seminar, the group opted to focus on the meta-level of the problem, developing a systematic approach to creating a comprehensive terminology mapping as a crowdsourced endeavor. This approach was then applied to a selected sample of SE fundamentals, enabling refinement of the methodology and providing a starting point for broader adoption by both communities.

In this article, we detail our systematic approach in \nameref{sec:approach}, including the term mapping schema, the initial identification of SE fundamentals, and the collaborative infrastructure developed to support this effort. Next, in \nameref{sec:results}, we present the preliminary findings derived from the application of this methodology to selected SE fundamentals extracted from the \emph{Guide to the Software Engineering Body of Knowledge}~\cite{SWEBOK3.0}. Specifically, we identified examples of (i) aligned SE fundamentals that are similarly understood and valued by both communities, (ii) SE fundamentals unfamiliar to the RSE community, presenting opportunities for education, training, and outreach to promote adoption by the RSE community, and (iii) SE fundamentals perceived as less useful by the RSE community, highlighting opportunities for adaptation and refinement by the SER community to make SE fundamentals more suitable for research software contexts. Finally, in \nameref{sec:plans}, we outline our plans for expanding this work by involving the broader RSE and SER communities. By gradually increasing engagement, we aim to develop a shared understanding and improve collaboration between these two communities, ultimately fostering better communication, mutual learning, and novel research directions.
\section{Approach}
\label{sec:approach}

Our approach leverages crowdsourcing as a means to gather input on terminology from both research software and software engineering research communities, with a website serving as the primary instrument to facilitate this process.  \addtext{Crowdsourcing has been effectively utilized in software engineering research as a means to leverage the collective intelligence and experience of a diverse group to tackle complex problems~\cite{Stol:2017:CSE}. Our platform features a discussion forum where practitioners can share their perspectives, fostering an environment for rich qualitative data collection. This interactive element addresses the need for dialogue and deeper understanding, as recommended for qualitative studies. It can also facilitate discussions among practitioners, allowing for the collection of qualitative data on terminology usage and perceptions. Our platform serves as an initial step toward building a comprehensive dataset of terminology used across communities. The insights gained will inform subsequent research phases, including interviews and repository mining, to explore the context and usage of specific terms more thoroughly.}

We have setup a schema to capture the mappings of individual terms, and we plan to engage multiple groups from both the RS and SE communities to flesh out the content.  In an attempt to be systematic, we are using the \emph{Guide to the Software Engineering Body of Knowledge}~\cite{SWEBOK3.0} as our primary source for SE terms.  In time, we expect to become more flexible in supporting multiple reference sources for SE terminology.  We are not aware of any systematic collection of RS terminology at this time, but we would be equally open to using such resources as they become available.  We then constructed a simple website, which is hosted on the GitHub Pages service, to share it with our target audience. 

\addtext{We believe this approach can serve as the basis for a useful research effort while requiring limited effort to maintain, which is an important consideration for an \textit{ad hoc} research activity without specific funding (and given the international nature of our team). The primary investment required in present work will be in the moderation and outreach efforts (see~\nameref{sec:governance} and~\nameref{sec:plans}), both of which are easily shared across volunteer participants.  At the same time, we believe that the data and understanding gained from this work can provide the basis and motivation for deeper and more focused research studies by interested community members that may warrant seeking separate funding.}

\subsection{Term Mapping Schema}
\label{sec:schema}


Listing~\ref{code:schema} presents the schema we developed to capture the essential term mapping information. Contributors complete a separate file for each term on the website.  Lines 1--7 pertain to the SE fundamental side of the mapping.  The \texttt{se\_fundamental} key (line 2) can take an array of values, allowing a set of synonyms or adjacent concepts to be specified.  The \texttt{fundamental\_description} (line 5) is a brief (approximately one sentence) description of the fundamental to make the displayed term mapping entry more readily understood.  The \texttt{swebok\_section} (line 8) connects the fundamentals to the section of the SWEBOK in which they are discussed.  This both facilitates display and navigation of the data (see~\nameref{sec:infrastructure}) and captures the connection to our source of ``ground truth'' for SE terminology.

Lines 10--14 capture similar information for the research software side of the mapping.  The \texttt{rse\_concept} (line 11) also allows an array of synonymous or adjacent terms to be specified.  \texttt{rse\_practice} (line 14) captures a brief description of typical realizations of the concept(s) in a research software setting.  This recognizes that the RS community may not interpret or implement SE fundamentals in exactly the same way as the SE community might envision.

\addtext{The fact that multiple values are supported for the \texttt{se\_fundamental}, \texttt{rse\_concept}, and \texttt{swebok\_section} keys provides a great deal of flexibility and avoids the limitations of 1:1 mappings.}

Lines 16--24 are meant to characterize the extent to which the SE fundamental is \emph{recognized} in the RS community (lines 18--19) and \emph{used} (lines 23--24).  The first entries (\texttt{rse\_awareness} and \texttt{rse\_usage}) are integers on a scale of 0--3, with 0 denoting effectively no awareness or usage of the SE fundamental in the RS community and 3 denoting widespread awareness or usage.  These values are intended to be rough ``t-shirt size'' characterizations rather than rigorously defined quantifications.  We felt that it would be useful to capture the  \emph{awareness} and \emph{usage} of a fundamental separately because they can tell us different things.  For example, an SE fundamental that is widely recognized in the RS community, but not widely used may suggest that the RS community has not found it useful for some reason, which might be worth further investigation (see below).  Similarly, if one felt that an SE fundamental would be really beneficial to the RS community, they might take different approaches to disseminating it depending on the level of recognition and usage observed.  The corresponding \texttt{*\_source} entries (lines 19 and 24) are intended to capture the source of the values for the awareness and usage.  Initially, we expect these will often be ``expert judgment'' or similar, implying a qualitative estimate by one or more practitioners.  However, we hope to be able to use surveys and other more quantitative mechanisms to gauge awareness and usage in the future, in which case the value might be ``survey'' and, ideally, a pointer to the published survey results would be provided under the \texttt{references} key (line 35).

Lines 26--32 capture similar assessments from the opposite perspective: what is the potential for software engineering research activities to improve the use of the fundamental in the RS community? The \texttt{ser\_potential} captures the rough magnitude of that potential, and \texttt{ser\_opportunities} provides the opportunity to capture a brief note about the nature of the SE research opportunity.

Rounding out the schema, \texttt{references} (line 35) allows a list (array) of links, papers and other material that may be useful to be captured, and \texttt{last\_reviewed} (line 38) captures the date on which the term entry was last reviewed or updated by the maintainers of the resource.  \addtext{At this stage, we are not defining a specific format for references, as we feel it is more important to have \emph{something} relevant rather than discourage adding references by making it overly prescriptive to supply them.  In many cases, a DOI link with a few words of context will likely suffice.  If we determine that more structure is needed in the references, there are several natural mechanisms available.  The Citation File Format\footnote{\url{https://citation-file-format.github.io/}} defined a YAML schema for citations which would be logical to incorporate into our schema, though we are not aware of existing integrations with Jekyll.  Another option may be BibTeX, which is another widely used structured format for representing citations which does have an integration with Jekyll.\footnote{\url{https://github.com/inukshuk/jekyll-scholar}}}

We expect that the schema may evolve as the term mapping effort proceeds -- both refining the core information already identified and extending the information we're collecting.

\begin{figure*}
\begin{lstlisting}[style=yaml, label={code:schema}, caption=Initial YAML schema for term mapping entry.]
# array of synonyms or adjacent concepts
se_fundamental:

# Short (1 sentence) description of the SE fundamental
fundamental_description:

# array of section identifiers
swebok_section: 

# array of synonyms or adjacent concepts
rse_concept:

# Text, a brief description of the typical realizations of the fundamental, in RSE practice
rse_practice: 

# General level of awareness of the fundamental in the research software community
# integers 0-3, 0=effectively no awareness, 3=widespread awareness
rse_awareness: 
rse_awareness_source:

# General level of usage of the fundamental in the research software community
# integers 0-3, 0=effectively no usage, 3=widespread use
rse_usage: 
rse_usage_source:

# Potential for SE research to improve use in research software
# integers 0-3, 0=effectively no opportunity, 3=significant SE research beneficial
ser_potential: 
ser_potential_source: 

# Reasons/opportunities for the SE research
ser_opportunities: 

# References (external links, papers, etc., that may provide useful connections)
references:

# Date of last review by the editorial board (YYYY-MM-DD)
last_reviewed: 
\end{lstlisting}
\end{figure*}

\subsection{Identification of Software Engineering Fundamentals}
\label{sec:fundamentals}


After initially experimenting with an ad hoc approach to identifying terms in either community, we decided that a more systematic approach would be useful.
For many years now, the SE community has developed and periodically updated the SWEBOK as a document that is intended to systematically capture the key aspects of the field of software engineering.
Since no large-scale efforts to capture a similar framework exists within the research software domain to our knowledge, we considered leveraging SWEBOK as a foundational reference for identifying initial terms to be a practical starting point.

Version 3 of the SWEBOK comprises 15 ``knowledge areas,'' each represented by a separate chapter:
\begin{enumerate}
    \item Software requirements
    \item Software design
    \item Software construction
    \item Software testing
    \item Software maintenance
    \item Software configuration management
    \item Software engineering management
    \item Software engineering process
    \item Software engineering models and methods
    \item Software quality
    \item Software engineering professional practice
    \item Software engineering economics
    \item Computing foundations
    \item Mathematical foundations
    \item Engineering foundations
\end{enumerate}
For the purpose of identifying SE terminology, concepts, practices, and tools (for which we use ``fundamentals'' as the generic term), we are currently focusing on the first ten knowledge areas in the SWEBOK, as we feel these are most directly connected to the practices research software developers are likely to be aware of and use.  However, we are open to reconsidering the importance of the remaining knowledge areas in helping to bridge between software engineering research and research software development.

SWEBOK chapters are further subdivided into sections and subsections, each discussing concepts within the knowledge area.  We are exercising our professional experience and judgment as SERs and RSEs ourselves to call out fundamentals of useful granularity for the mapping initiative.  Not surprisingly, in many cases, we find that a SWEBOK subsection offers an appropriate level, while in other cases, we may consolidate several subsections or treat an entire section as a term for the mapping exercise.  For example, SWEBOK may differentiate concepts that (in our experience) RS developers do not.  Of course, we can easily break up such entries into multiple terms if feedback from users of the website indicates that we have misjudged.

\subsection{Infrastructure for Collaboration}
\label{sec:infrastructure}


In order to share the evolving results for the term mapping effort with the public, and to solicit input, we used the Jekyll static website generator\footnote{\url{https://jekyllrb.com/}} and GitHub Pages\footnote{\url{https://pages.github.com/}} to create a sub-site\footnote{\url{https://ser-rse-bridge.github.io/mapping-of-terms/}} of the website for our Dagstuhl Seminar.\footnote{\url{https://ser-rse-bridge.github.io/}}. 

The main page of the term mapping site explains the goals and approach and presents the table of contents (TOC) of the SWEBOK, which includes the section numbers, headings, page numbers, and a link to the mapping, if available. Given that we are relying on the SWEBOK to guide us systematically through SE terminology, it seemed logical to also use it as the primary entry point to the content of the site.  Even for visitors who are unfamiliar with SWEBOK and its structure, we believe that the systematic and hierarchical approach is easily understood and navigated.  

To produce the SWEBOK TOC for the Jekyll site, we extracted the text of the table of contents from the PDF version of the SWEBOK and processed it into a tab-separated variable (TSV) file containing the fields:
\begin{enumerate}
    \item Section number
    \item Section heading
    \item Page number
    \item Whether the section contains terms that should be mapped
\end{enumerate}
The section numbers in the TSV file have been modified from their appearance in the original SWEBOK document to facilitate automatic processing in the generation of the website.  Specifically, at each level, they have been zero-filled to two digits, and are represented as one, two, or three two-digit numbers, depending on the heading level, separated by periods (``.'').
The fourth field was populated manually, in conjunction with the identification of the SE fundamental, discussed above.  It contains either ``n/a'', indicating that the section has no applicable terms to map, or it is blank, indicating that we expect to have terms from this section.
\addtext{Since all levels of the SWEBOK document hierarchy are available in the TOC file, we can easily associate the term files with entries at whatever level is appropriate, as discussed above.}

For those who are familiar with the SWEBOK, it seemed most natural to display the \emph{entire} table of contents, including sections which have no applicable terms to map.  Therefore, in presenting the SWEBOK TOC, we display sections without applicable terms to map in red and denote ``n/a'' in the fourth column.  As the TSV file is processed, we scan the current collection of terms to determine whether or not we have a mapping for that section yet and display the link to the mapping or leave the field blank, accordingly.

For the individual terms, we use the Jekyll ``collections'' feature.  Each term is represented as a separate file in the source for the site.  The files are standard Jekyll Markdown format, which includes a YAML ``frontmatter'' section.  The primary content of a term entry is in the YAML frontmatter, which should conform to the schema presented in \addtext{Listing}~\ref{code:schema}.  The body of the file may contain any further notes contributors may wish to add, using Markdown markup.  The rendering of the term entries is done with the standard Jekyll approach, which uses template ``layouts'' and code in the Liquid scripting language.\footnote{\url{https://shopify.github.io/liquid/}}

Users of the site may leave comments on the main page or any of the term pages.  We anticipate comments as being a primary means for visitors to contribute to the content of the site.  The Minimal Mistakes\footnote{\url{https://mmistakes.github.io/minimal-mistakes/}} Jekyll theme we chose to use for the site has built-in support for comments with multiple backends.  We chose the giscus comment system,\footnote{\url{https://giscus.app/}} which is based on GitHub's Discussions capability.  

Based on these implementation choices, the entire term mapping site, including content, code, and comments, is conveniently contained within a single GitHub repository\footnote{\url{https://github.com/ser-rse-bridge/mapping-of-terms}} and the site is served through GitHub's Pages service.  Our hope is that our target audience, developers of research software and software engineering researchers, will find the implementation straightforward and transparent, and will be comfortable contributing through raising issues and contributing pull requests as well as through comments. An additional benefit of using the GitHub infrastructure is that it naturally provides a version-controlled history of all the changes that are made in the evolution of the website. Most importantly, this includes the history of the mapping of the terms themselves.

\subsection{\addtext{Process and Governance}}
\label{sec:governance}

\addtext{As described in \nameref{sec:plans}, we plan to recruit a successively larger group of contributors to the site.  We expect that the primary form of input from the community will be through the comment mechanism integrated into the site. Following the spirit of open peer review, comments will be available for all to see.  We plan to assemble groups for synchronous online or in-person discussions as well.  Such discussions will also be summarized through comments, so that they're available to all on the same platform.  We will establish a code of conduct for contributors to the site, and the editors will moderate discussions as necessary to ensure civility and guide them to stay on topic. (The GitHub Discussions mechanism we are leveraging for the site's comment capability also allows discussions that aren't specifically associated with term mappings.)  The members of our working group will comprise the initial editorial board for the site, but it can be expanded as needed.}

\addtext{When there is a consensus from the discussions that changes are needed to the term mapping entry, they will be made using the pull request process of the GitHub hosting environment. We expect that most such pull requests will be initiated by members of the editorial board, but the process is open to community participants as well.  Pull requests will be reviewed and approved by several editorial board members (specifics to be determined) before they are merged into the site and noted in the accompanying comments.  The site's git repository and pull request tracking tools ensure that the complete history of changes is available for inspection.}

\addtext{As with any community, we expect that there will sometimes be differences of opinion.  Where necessary, we expect the editorial board to resolve such differences, perhaps with the assistance of a small group of subject matter experts where the board deems it helpful. Again, we expect to follow the spirit of open peer review and document the discussions and decisions as part of the record.  Given the nature and goals of the site we do not expect the differences to be particularly contentious, nor the consequences of decisions that may be made to have adverse impacts on the participants or the community.  Points of disagreement within the community are also likely to be good targets for more detailed investigations by interested researchers, perhaps using more quantitative methodologies.}

\subsection{\addtext{Future Evolution}}
\label{sec:evolution}

\addtext{We are taking an agile approach to the design and development of the term mapping schema and website.  As noted, we expect to revise our approach as we gain more experience.  This may involve changes to the schema as well as the way is it presented on the site.  We consider the flexibility of the overall Jekyll-based approach we have adopted to be a benefit in this regard.  Probably one of the more significant changes that we can anticipate in a general sense is the addition of additional sources of ``ground truth'' for the definition and differentiation of terms.  We've started with the SWEBOK as our sole reference, but as noted above, we anticipate adding other sources as we feel that we've covered SWEBOK and as we identify other sources that would be recognized as canonical.  In terms of the schema, this may be as simple as replacing the \texttt{swebok\_section} key with a set of keys that allow specification of both the reference document and the section within that document. However, from a presentation and governance perspective, we will have to determine whether terms a limited to a single source per page, or whether we want to be able to reference (equivalent) terms within multiple sources in a single page, and how to handle possible differences in their definition and usage between different sources.  We feel that these decisions will have to be based on the specifics of the additional references chosen and therefore must be deferred until the situation comes up. A related change that may need to be made even sooner is for a situation where someone identifies a term in use in the RSE community that does not appear to have an equivalent in the SER community, and we want to create a mapping file for it to solicit community input and discussion. While the schema would support this case, the presentation of the table of contents level on the website would need to be revised, and the considerations would be essentially the same as introducing a second source of ground truth on the SE side.}
\section{Emerging Results}
\label{sec:results}

The initial ad hoc discussions of the working group looked at the software life cycle and the various processes and tools used by both communities.
For example, both communities value version control as the central tool to document and manage the code development process.
Both groups also use the term ``technical debt'' in the same manner, although the SE researchers also use the term ``code smells'', which is not widely used in the RS community.

After our thinking evolved into the more structured term mapping approach, the working group adapted the previously identified commonalities and discrepancies in term mapping into the presented structure.
Furthermore, the group also took the newly established SWEBOK-based fundamentals and validated the proposed method by adding further exemplary entries to the collection of term mappings.
These initial contributions highlighted three types of outcomes: an alignment between terms used in both communities, a lack of awareness on the RSE side or a lack of adoption by RSEs.

\subsection{Alignment of Terms}

For a number of terms an equivalent practice could be identified in the short time of the Seminar.
However, it also became also obvious that the two communities can use (slightly) different terms for the same concept.
For example, while SWEBOK uses the distinct terms 
``requirement elicitation'' (SWEBOK section 01.03), and ``requirement analysis'' (SWEBOK section 01.04), RSEs typically use ``requirements gathering'' and ``requirements analysis''  for both concepts combined.
In our discussions, we concluded that techniques such as ``user stories'' and ``list of requirements'' are used similarly in both communities, with apparent differences stemming more from the quality level of the software to be developed.

It is these examples of comparable techniques that also motivate the overall approach of mapping (research) software engineering terms, as this provides the easiest ``bridge'' between the communities: it is just a translation of terms.
As a result, also in the discussions at the seminar the group could dive very quickly into technical discussions about these fundamentals and quickly build an understanding of what both sides can learn from each other.

\subsection{Lack of Awareness}

Our initial discussions also highlighted areas where RSEs lack awareness of SE fundamentals.
For example, while SWEBOK has a whole section on testing, distinguishing numerous different \addtext{types of tests} (section 04), many developers of research software include only a few sample input/output tests to test their software without making the distinctions the SWEBOK makes.
The feeling of the RSEs in our working group is that many developers of RS would consider the numerous different types of testing defined by the SER community to be both overwhelming, and excessive, given that a great deal of research software is primarily developed as a limited-lifetime prototype or demonstration.
At the same time, RSEs are generally aware that they are accruing technical debt due to not doing ``enough'' testing and fully acknowledge the need for better and more structured testing -- in a general sense.
Leveraging the experience of software engineering researchers on how they utilize different types of testing and integrate them into software development processes with low overhead or other resource constraints is an opportunity to enhance RSE practice.

\subsection{Lack of Adoption}

One surprising outcome of the initial discussions of our working group was that the opposite can also happen.
While both sides are---as previously mentioned---aware of requirement analysis, the RSEs in the group explained that the very structured and details approach outlined in SWEBOK often does not hold up in their daily work beyond simple user stories or short item lists of requirements.
The discussion of the ``why'' revealed that research software is often an integral part of the research process, which is described as non-linear or more exploratory in its nature.
Hence, the requirements for the software also often change in these prototyping phases.
On the other hand, for some types of research software, e.g., that which is part of a large research infrastructure, such particle accelerators, microscopes or other instrumentation, the software requirements are often directly derived from the infrastructure itself.

Identifying gaps in adoption of SE fundamentals by the RS community provides opportunities to explore the reasons for the gap and whether the perceptions RSEs have of the fundamental are justified or, perhaps, represent a misunderstanding of the SER practice or intent.  Depending on the outcomes, such explorations could lead to improvements in RSE practice or to an expanded understanding of the software engineering landscape by SERs.



\section{Future Plans}
\label{sec:plans}


To ensure the impact of our mapping initiative, we plan to broaden our pool of contributors significantly. 
Starting with the participants of the Dagstuhl Seminar, we will ask them to contribute to a ``first pass'' through the current state, providing a baseline position against which to compare and tension future contributions. We shall then invite their teams and close collaborators to contribute additions and changes to the mappings. Here we will iteratively review the progress within our working group, ensuring coverage and consistency. 
Finally, we shall ask the wider RSE and SER communities to contribute. 

From the RSE side, we will initially target established communities such as the Society of Research Software Engineering,\footnote{\url{https://society-rse.org/}} US-RSE\footnote{\url{https://us-rse.org/}}, Digital Research Alliance of Canada\footnote{\url{https://alliancecan.ca/en/services/research-software}}, and de-RSE.\footnote{\url{https://de-rse.org}}
These organizations not only provide vital resources and support for RS practitioners but also serve as hubs for networking, knowledge exchange, and the dissemination of best practices. By collaborating with these groups, we aim to leverage their collective expertise and reach to maximize the impact of our efforts.

From the SER side, we will focus on fostering connections with academic and industrial communities engaged in the systematic study of software engineering methods, tools, and processes. This includes outreach to prominent conferences (e.g., the International Conference on Software Engineering) as well as national and international societies and organizations (e.g., ACM SIGSOFT\footnote{\url{https://www2.sigsoft.org}}, the IEEE Computer Society\footnote{\url{https://www.computer.org}}, or relevant IFIP Technical Committees and Working Groups\footnote{\url{https://dl.ifip.org/ifip.html}}, to ensure that our initiatives are grounded in commonly accepted knowledge and aligned with the evolving priorities of the SER community.  
After our community contribution period, we will re-validate the current state of the mappings by asking Dagstuhl Seminar participants to review them, following up with interviews to resolve disagreements and ambiguities, or where serious concerns are raised. This will allow us to determine the perceived quality and usefulness of the submissions. In sum, we believe our endeavor to be a significant stepping stone towards a shared understanding that fosters collaboration, drives innovation, and enhances the quality of research software engineering and unlocks new directions for software engineering research.


Further validation of the mappings from SE to RSE terms will be performed by starting from the RSE terms and mapping them to SE terms. We anticipate that the majority of terms already mapped will be (at least approximately) commutative---that is map back from RSE to SE---but there may be some subtleties that are useful to capture, and we expect that there will be terms used in the RSE community that have not been captured in the mapping from SE to RSE.

\section{ACKNOWLEDGMENTS}

This material is based in part upon work supported by the U.S. Department of Energy, Office of Science, Office of Advanced Scientific Computing Research, Next-Generation Scientific Software Technologies program, under contract number DE-AC05-00OR22725 to ORNL.

\bibliographystyle{IEEEtran}
\bibliography{main}


\begin{IEEEbiography}{Timo Kehrer} is a permanent professor at the Institute of Computer Science at the University of Bern, Switzerland, chairing the Software Engineering research and teaching Group (SEG). He is the current president of CHOOSE, the Swiss Group for Original and Outside-the-box Software Engineering, a special interest group of the SI (Swiss Informatics Society). Kehrer has broad interests in various areas of software engineering research, and he is particularly interested in bridging the gap between Research Software Engineering and Software Engineering Research through his association with the Collaborative Research Center FONDA (Foundations of Workflows for Large-Scale Scientific Data Analysis).
\end{IEEEbiography}

\begin{IEEEbiography}{Robert Haines} is Director of Research IT and an Honorary Lecturer at the University of Manchester, and a Fellow of the Software Sustainability Institute. He is one of the originators of the term ``Research Software Engineer'', served for six years as an elected representative of the UK RSE Association, chaired the First Conference of Research Software Engineers in 2016, and was a founding trustee of the Society of Research Software Engineering. Robert's research interests include software engineering, software sustainability, software use in open and reproducible research, software citation and credit, and career paths for software engineers and data scientists.
\end{IEEEbiography}

\begin{IEEEbiography}
{Guido Juckeland} is the head of the Computational Science Department at Helmholtz-Zentrum Dresden Rossendorf. His research interests include increasing scientific productivity and reproducibility through better software and data management processes, tooling, training and recognition of this type of work.
\end{IEEEbiography}

\begin{IEEEbiography}
{Shurui Zhou} is an assistant professor at the University of Toronto. Her research focuses on helping distributed and interdisciplinary software teams to collaborate more efficiently and build high-quality software systems, especially in the context of modern open-source collaboration forms, fork-based development, and interdisciplinary teams when building AI-enabled systems or scientific software. 
\end{IEEEbiography}

\begin{IEEEbiography}{David E.~Bernholdt} is a distinguished R\&D staff member at Oak Ridge National Laboratory.  His research interests focus on the development of scientific software for high-performance computers, including developer productivity, and software quality and sustainability.
\end{IEEEbiography}

\end{document}